\documentclass[prb,showpacs,preprintnumbers,amsmath,amssymb,floatfix,twocolumn,superscriptaddress]{revtex4}

\usepackage[english]{babel}
\usepackage{graphicx}     
\usepackage{dcolumn}     
\usepackage{bm}               
\usepackage{color}
\usepackage[colorlinks=true ,citecolor=blue]{hyperref}

\newcommand{\ee}{\text{e}}
\newcommand{\ii}{\text{i}}
\newcommand{\so}{\text{SO}}
\newcommand{\nn}{\text{N}}

\newcommand{\kets}[1]{|\!#1\rangle}

\newcommand{\bra}[1]{\langle\,#1|}
\newcommand{\ket}[1]{|#1\,\rangle}

\newcommand{\comment}[1]{}

\date{\today}

\begin{document}

\title{Spin-resolved scattering through spin-orbit nanostructures in graphene}
\author{D. Bercioux}
\email{dario.bercioux@frias.uni-freiburg.de}
\affiliation{Freiburg Institute for Advanced Studies, Albert-Ludwigs-Universit\"at, D-79104 Freiburg, Germany}
\affiliation{Physikalisches Institut, Albert-Ludwigs-Universit\"at, D-79104 Freiburg, Germany}
\author{A. De Martino}
\email{ademarti@thp.uni-koeln.de}
\affiliation{Institut f\"ur  Theoretische Physik, Universit\"at zu K\"oln, Z\"ulpicher Stra\ss e 77, D-50937 K\"oln, Germany}

\begin{abstract}
We address the problem of spin-resolved scattering through 
spin-orbit nanostructures  in graphene, \emph{i.e.}, regions of 
inhomogeneous spin-orbit coupling on the nanometer scale. 
We discuss the phenomenon of spin-double refraction and its consequences
on the spin polarization. Specifically, we study the transmission properties of a 
single and a double interface between a normal region and a region with finite 
spin-orbit coupling, and analyze the polarization properties of these systems. 
Moreover, for the case of a single interface, we determine the spectrum of 
edge states localized  at the boundary between the two
regions and study their properties.
\end{abstract}



\maketitle

\section{introduction}
\label{intro}

Graphene\cite{geim,kim} | a single layer of carbon atoms arranged in a 
honeycomb lattice | has attracted  huge attention in the physics community
 because of many unusual electronic, thermal and nanomechanical properties.\cite{reviews,beenakker:2008}
In graphene the Fermi surface, at the 
charge neutrality point, reduces to two isolated
points, the two inequivalent 
corners $K$ and $K'$ of the hexagonal Brillouin zone of the honeycomb lattice. 
In their vicinity the charge carriers form a gas of chiral 
massless quasiparticles with a characteristic conical spectrum. 
The low-energy dynamics is governed by the Dirac-Weyl (DW)
equation\cite{semenoff,divincenzomele} in which the role of speed of light 
is played by the electron Fermi velocity. 
The chiral nature of the quasiparticles 
and their linear spectrum lead 
to remarkable consequences  for a variety of electronic properties as weak 
localization, shot noise, Andreev reflection, and many others. 
Also the behavior in a perpendicular magnetic field discloses 
new physics.  Graphene exhibits a zero-energy Landau level, 
whose existence gives rise to an unconventional half-integer quantum Hall effect,
one of the peculiar hallmarks of the DW physics. 

Driven by the prospects
of using this material in spintronic applications,\cite{spintronics,Trauzettel:2008}  the study of 
spin transport is one of the most active field in graphene research.\cite{expspin1,expspin2,expspin3,expspin4,expspin5,expspin6}
Several experiments  have recently demonstrated 
spin injection, spin-valve effect, and spin-coherent transport in graphene, 
with spin relaxation length of the order of few micrometers.\cite{expspin2,
expspin6} In this context a crucial role is played by the spin-orbit interaction. 
In graphene symmetries allow for
two kinds of spin-orbit coupling (SOC).\cite{kane:2005}
The {\em intrinsic} SOC originates 
from carbon intra-atomic SOC. It opens a gap in the energy spectrum 
and converts graphene into a topological insulator 
with a quantized spin-Hall effect.\cite{kane:2005}
This term has been estimated to be rather weak in clean flat 
graphene.\cite{soguinea,so2,so3,so4} 
The {\em  extrinsic} Rashba-like SOC originates instead from interactions 
with the substrate, presence of a perpendicular external
electric field, or curvature of graphene membrane.\cite{soguinea,so2,so3,so5} 
This term is believed to be responsible for 
spin polarization\cite{rashbagraphene}
and spin relaxation\cite{guinea2009,ertler:2009} physics in graphene.
Optical-conductivity 
measurements could provide a way to determine the respective
strength of both SOCs.\cite{philip}

In this article we address the problem of ballistic 
spin-dependent 
scattering in the presence of
inhomogeneous spin-orbit couplings.
Our main motivation stems from a recent experiment 
that reported a large enhancement of Rashba SOC splitting
in  single-layer graphene grown on Ni(111) intercalated with a 
Au monolayer.\cite{varykhalov:2008} 
Further experimental results show that the intercalation of Au atoms 
between graphene and the Ni substrate is essential in order to 
observe sizable Rashba effect.\cite{rader:2009,varykhalov:2009}
The preparation technique of Ref.~\onlinecite{varykhalov:2008}
seems to provide a system with properties very close 
to those of  freestanding graphene in spite of the fact that 
graphene is grown on a solid substrate. The presence 
of the substrate does not seem to fundamentally alter 
the electronic properties observed in suspended systems, 
\emph{i.e.},  the existence of Dirac points at the Fermi energy and
the gapless conical dispersion in their vicinity.

These results suggest that a certain degree of control 
on the SOC can be achieved by appropriate substrate engineering, 
with variations of the SOC strength on sub-micrometer scales,
without spoiling the relativistic gapless nature of quasiparticles.
This could pave the way for the realization of spin-optics devices 
for spin filtration and spin control for DW fermions in graphene. 
An optimal design would  require a detailed understanding of the spin-resolved ballistic
scattering through such \emph{spin-orbit nanostructures}, which is the aim
of this paper.

The problem of spin transport through nanostructures with inhomogeneous SOC
has already been thoroughly studied in the case of two-dimensional electron 
gas in semiconductor heterostructures with Rashba 
SOC.\cite{marigliano1, khodas,marigliano2}
Here the Rashba SOC~\cite{rashba:1960} | arising from the
inversion asymmetry of the confinement potential |  
couples the electron momentum  to the spin degree of freedom 
and thereby lifts the spin degeneracy. 
In this case, a region with finite SOC 
between two normal regions has properties similar to biaxial 
crystals: an electron wave incident from the normal region 
splits at the interface and the two resulting waves propagate in the SO region 
with different Fermi velocities and momenta.~\cite{marigliano1} 
This effect | analogous to the optical  double-refraction | 
produces an interference pattern when the electron waves emerge in the second 
normal region. Moreover, electrons that are injected in an spin unpolarized state 
emerge from the SO region in a partially polarized state.

Here we shall focus on the two simplest examples of 
SO nanostructures in graphene:  
(i) a single interface between two regions with different strengths of SOC;
(ii) a SOC barrier, or double interface, \emph{i.e.}, a region of finite SOC in between
two regions with vanishing SOC.

Our analysis shows | in analogy to the case of 2DEG |  that
the ballistic propagation of carriers
is governed by the spin-double refraction. 
We find that the scattering
properties of the structure 
strongly depend on the injection angle. As a consequence, 
an initially unpolarized DW quasiparticle emerges from the SOC barrier 
with a finite spin polarization.  
In analogy to the edge states in the quantum spin-Hall 
effect,\cite{kane:2005} we also consider the possibility 
of edge states localized at the interface between regions with and without SOC.

This paper is organized as follows. In Sec.~\ref{sec:2} we introduce
the model and the transfer matrix formalism used in the rest of the paper.
In Sec.~\ref{sec:3} we discuss the scattering problem at a single interface 
 and the spectrum of edge states.  In Sec.~\ref{NSON} we address 
the case of a double interface | a SOC barrier | and the final Sec.~\ref{sec:5} 
is devoted to the discussion of results and conclusions.

\section{Model and Formalism}
\label{sec:2}

%
%
\begin{figure}[!t]
	\centering
	\includegraphics[width=\columnwidth]{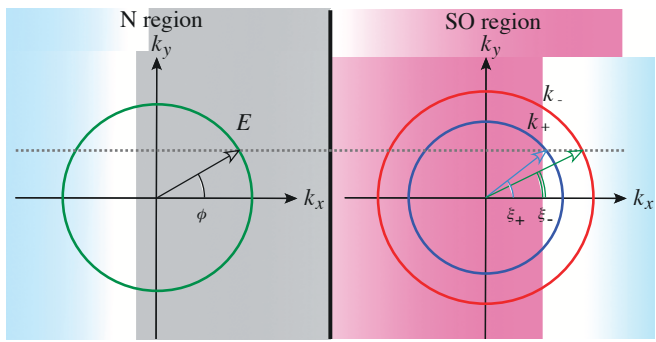}
	\caption{\label{fig:kinetics} (Color online) Illustration of the kinematics of the 
	scattering at a N-SO interface in graphene. The circles represent
	constant energy contours.}
\end{figure}
%
%

We consider a clean graphene sheet
in the $xy$-plane with SOCs\cite{kane:2005,soguinea,
rashbagraphene,stauber:2009,yamakage:2009} inhomogeneous 
along the $x$-direction. We shall restrict ourselves
to a single-particle picture and neglect electron-electron interaction effects.
The length scale over which the SOCs vary 
is assumed to be much larger than graphene's lattice constant
($a=0.246$~nm) but much smaller than the typical 
Fermi wavelength of quasiparticles $\lambda_\text{F}$. 
Since close to the Dirac points $\lambda_\text{F} \sim 1 / |E|$, 
at low energy $E$ this approximation is justified. 
This assumption ensures that we can use the 
continuum DW description, in which the two valleys are not coupled.
Yet close to a Dirac point we can approximate the variation 
of SOCs as a sharp change.
Focusing on a single valley, the single-particle Hamiltonian reads
%
%
\begin{align}\label{eq:one}
\mathcal{H}  & = v_\text{F}\ {\bm \sigma} \cdot  {\bf p} + \mathcal{H}_\so,\\
\mathcal{H}_\so & = \frac{\lambda(x)}{2} \left(
{\bm \sigma} \times {\bf s}\right)_z + \Delta(x) \sigma_z s_z,\label{eq:ham:so}
\end{align}
%
%
where $v_\text{F}\approx 10^6$ m/s is the Fermi 
velocity in graphene. In the following we set $\hbar =v_\text{F}=1$.
The vector of Pauli matrices $\bm \sigma=(\sigma_x,\sigma_y)$ 
[resp. ${\bf s} =(s_x,s_y)$] 
acts in sublattice space [resp. spin space]. 
The term $\mathcal{H}_\so$ contains the extrinsic or 
Rashba SOC of strength $\lambda$ 
and the intrinsic SOC of strength $\Delta$. 
While experimentally  the Rashba SOC can be enhanced 
by appropriate optimization of the substrate up to values of the order
of $14$ meV,~\cite{varykhalov:2008} the intrinsic SOC
seems at least two orders of magnitude smaller. 
Yet, the limit of large intrinsic SOC is of considerable interest
since in this regime graphene becomes a topological insulator.\cite{kane:2005} 
Thus in this paper we shall not restrict ourself to the 
experimentally relevant  regime $\lambda\gg \Delta$ but consider 
also the complementary regime.

The wave function $\Psi$ is expressed as 
%
%
\begin{equation*}
\Psi^\text{T}=(\Psi_{A\uparrow},\Psi_{B\uparrow},
\Psi_{A\downarrow},\Psi_{B\downarrow}),
\end{equation*}
%
%
where the superscript ${}^\text{T}$ denotes transposition.
Spectrum and eigenspinors of the Hamiltonian (\ref{eq:one}) with uniform
SOCs are briefly reviewed in Appendix \ref{app:so}. 
The spectrum consists of
four branches $E_{\alpha,\epsilon}({\bf k})$ labelled by 
the two quantum numbers $\epsilon=\pm1$ and $\alpha=\pm1$.
Here, the first distinguishes particle and hole branches, 
the second gives the sign of the expectation value of the 
spin projection along the in-plane direction
perpendicular to the propagation direction ${\bf k}$.
The spectrum strongly depends on the ratio 
%
%
\begin{equation}\label{eq:eta}
\eta=\frac{\Delta}{\lambda}. 
\end{equation}
%
%
For $\eta>1/2$ a gap 
separates particle and hole branches. The gap closes at $\eta=1/2$ 
and for $\eta<1/2$ one particle branch and 
one hole branch are 
degenerate at ${\bf k}=0$ (see Fig.~\ref{fig:spec} in App.~\ref{app:so}).

We now briefly summarize the transfer matrix approach
employed in this paper to solve the DW scattering 
problem.~\cite{mckellar:1987,peeters,luca1,luca2} 
We assume translational invariance in the $y$-direction, thus 
the scattering problem for the Hamiltonian (\ref{eq:one})
reduces 
to an effectively one-dimensional (1D) one.
The wave function factorizes as $\Psi(x,y)=\ee^{\ii k_yy}\chi(x)$, where $k_y$
is the conserved $y$-component of the momentum, which parameterizes
the eigenfunctions of the Hamiltonian of given energy $E$.

For simplicity we consider piecewise constant profiles of SOCs, 
and solve the DW equation in each region 
of constant couplings.
Then we introduce the $x$-dependent $4\times 4$ matrix $\Omega(x)$,  whose 
columns are given by the components of the independent, normalized
eigenspinor of the 1D DW Hamiltonian at fixed energy.\cite{footnote1}
Due to the continuity  of the wave function at each interface between 
regions of different  SOC, the wave function on the left of the interface 
can be expressed in terms of  the wave function on the right via the 
transfer matrix
%
%
\begin{equation}\label{eq:tran:mat}
\mathcal{M}= \left[\Omega(x_0^-)\right]^{-1}\Omega(x_0^+),
\end{equation}
%
%
where $x_0$ is the position of the interface and $x_0^\pm=x_0\pm \delta$
with infinitesimal positive $\delta$. 
The condition $\det \mathcal{M}=1$ guarantees 
conservation of the probability current across the interface. The generalization 
to the case of a sequence of $N$ interfaces at positions $x_i$, $i=1,\dots,N$, 
is straightforward since the transfer matrices relative
to individual interfaces combine 
via matrix multiplication: 
%
%
\begin{equation}\label{eq:tran:mat:2}
\mathcal{M} = \prod_{i=1}^N  \left[\Omega(x_i^-)\right]^{-1}\Omega(x_i^+).
\end{equation}
From the transfer matrix  it is straightforward to determine 
transmission and reflection matrices, which encode all the 
relevant information on the scattering properties.


\section{The N-SO interface}
\label{sec:3}

%
%
\begin{figure}[!tb]
	\begin{center}
	\includegraphics[width=\columnwidth]{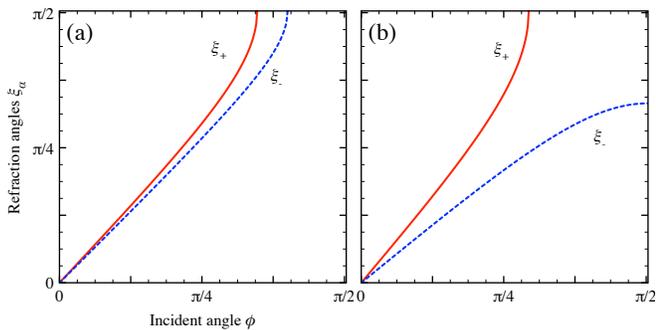}
	\caption{\label{fig:angles} (Color online) Refraction angles 
	as function of the incidence angle for fixed energy
	and fixed SOCs.  
	Panel (a): $E=5$, $\lambda=0.5$, $\Delta=2$;
	panel (b): $E=5$, $\lambda=2$, $\Delta=0.5$.}
	\end{center}
\end{figure}
%
%

First we concentrate on the elastic scattering problem at the 
interface separating a normal region N ($x<0$), where SOCs vanish,  
and a SO region ($x>0$), where SOCs
are finite and uniform.

We consider a quasiparticle of energy $E$, with $E$ assumed positive for 
definiteness and outside the gap possibly opened by SOCs.
This quasiparticle incident from the normal region  
is characterized by the $y$-component of the momentum, or equivalently,
the incidence angle $\phi$ measured with respect to the normal at the 
interface, see Fig.~\ref{fig:kinetics}.
Conservation of $k_y$ implies that
%
%
\begin{subequations}
\begin{align}
k_y^\nn   &= E \sin \phi  = E_{\alpha}\sin \xi_\alpha = k_y^\so \label{eq:mc} \\
k_x^\nn & =  E \cos \phi \label{eq:mc:a}\\ 
k_{x\alpha}^\so & = E_{\alpha} \cos\xi_{\alpha}\label{eq:mc:b}
\end{align}
\end{subequations}
%
%
where $\alpha=\pm 1$ and $E_\alpha=\sqrt{(E-\Delta)(E+\Delta-\alpha\lambda)}$. 
The refraction angles $\xi_\alpha$ are 
fixed by momentum conservation along the interface (\ref{eq:mc}) and read
%
%
\begin{equation}\label{eq:ref:angles}
\xi_\alpha = \arcsin\left( \frac{E}{E_\alpha}\sin\phi \right)\,.
\end{equation}
%
%
Figure~\ref{fig:kinetics} illustrates the refraction process at the N-SO interface. 
The incident wave function, assumed to have fixed spin projection in the 
$z$-direction, in the SO region splits in a superposition of eigenstates of the 
SOCs Hamiltonian corresponding to states in the different branches of the spectrum.  
These eigenstates propagate along two distinct directions characterized by 
the angles $\xi_\alpha$, whose difference depends on SOC and is an 
increasing function of the incidence angle, 
see Fig.~\ref{fig:angles}.
The angles $\xi_\alpha$ coincide only  for normal incidence or for  $\lambda=0$. 

Equation (\ref{eq:ref:angles}) implies that there exists a critical angle for 
each of the two modes given by
%
%
\begin{equation}\label{eq:crit:angles}
\tilde\phi_\alpha = \arcsin\left( \frac{E_\alpha}{E} \right).
\end{equation}
%
%
For $\phi$ larger than both critical angles $\tilde \phi_\alpha$, the
quasiparticle is fully reflected, since there are no available transmission 
channels in the SO region.
For $\phi$ in between the two critical angles  
the quasiparticle transmits only  in one channel.\cite{footnote2} 

After this qualitative discussion of the
kinematics,  we now present the
exact solution of the scattering problem. 
In the N region  $x<0$ a normalized scattering state of energy $E>0$, 
incident from the left on the interface with incidence angle $\phi$  
and  spin projection $s$ is given by 
%
%
\begin{align}\label{eq:wf:norm}
\chi_\nn(x)  = & 
\left[ \delta_{\uparrow,s} \kets{\uparrow} + 
\delta_{\downarrow,s}  \kets{\downarrow}\right] 
\begin{pmatrix}
1\\ \ee^{\ii \phi}
\end{pmatrix}
\frac{\ee^{\ii k_x x}}{\sqrt{ 2v_\text{F}^x}}  
 \nonumber\\
& + \left[ r_{\uparrow s} \kets{\uparrow}+ r_{\downarrow s} \kets{\downarrow} \right]
\begin{pmatrix}
1\\ -\ee^{-\ii\phi}
\end{pmatrix}
\frac{\ee^{-\ii k_x x}}{\sqrt{ 2v_\text{F}^x}},
\end{align}
%
%
where $k_x\equiv k_x^\nn$ (cf. Eq.~\ref{eq:mc:a}). Here
the index $s=\uparrow,\downarrow$ specifies the 
spin projection  of the incoming quasiparticle with 
$\kets{\!\uparrow}$ and $\kets{\!\downarrow}$ eigenstates of $s_z$
and $\delta_{i,j}$ is the Kronecker delta.
The velocity $v_\text{F}^x=\cos\phi$ is included to ensure 
proper normalization of the scattering state.
The complex coefficients $r_{s' s}$
are reflection probability amplitudes for a 
quasiparticle with spin $s$ to be reflected with spin $s'$.
The associated  matrix $\Omega_\text{N}(x)$ reads
%
%
\begin{align}
\Omega_\nn(x)
& =  \frac{1}{\sqrt{ 2v_\text{F}^x}} \nonumber \\ &
 \begin{pmatrix}
 \ee^{\ii k_x x} & \ee^{-\ii k_x x}  & 0 & 0\\
 \ee^{\ii (k_x x+\phi)} & -\ee^{-\ii (k_x x+\phi)} & 0 & 0\\
0 & 0 &\ee^{\ii k_x x} & \ee^{-\ii k_x x} \\
0 & 0 &\ee^{\ii (k_x x+\phi)} & -\ee^{-\ii (k_x x+\phi)} \nonumber
\end{pmatrix}.
\end{align}
%
%
Similarly the wave function in the SO region ($x>0$) 
can be expressed in general form as
%
%
\begin{align}\label{eq:wf:so}
\chi_\so(x) & =  \frac{1}{\sqrt{v_{++}^x}}\left[t_+\psi_{++}(x) 
 + r_+\bar\psi_{++}(x)\right] \nonumber \\ &
  + \frac{1}{\sqrt{v_{-+}^x}}\left[t_-\psi_{-+}(x) + r_-\bar\psi_{-+}(x)\right]
\end{align}
%
%
where $t_\pm$ (resp. $r_\pm$) are complex amplitudes for right-moving 
(resp. left-moving) states. The coefficient $t_\alpha$ represents
the transmission amplitude into mode $\alpha$.
The wave functions $\psi_{\alpha+}$ and the 
Fermi velocities $v_{\alpha+}^x$ in the SO region are obtained from
the expressions given in App.~\ref{app:so} with the replacement
$k_x\to k_{x\alpha}^\so$, where for notational 
simplicity the label $\so$ will be understood.
The wave functions $\bar\psi_{\alpha+}$
are  in turn obtained from $\psi_{\alpha+}$ by replacing  
$k_{x\alpha} \to -k_{x\alpha}$. 
The matrix $\Omega_\so(x)$ then reads 
%
%
\begin{figure}
	\includegraphics[width=\columnwidth]{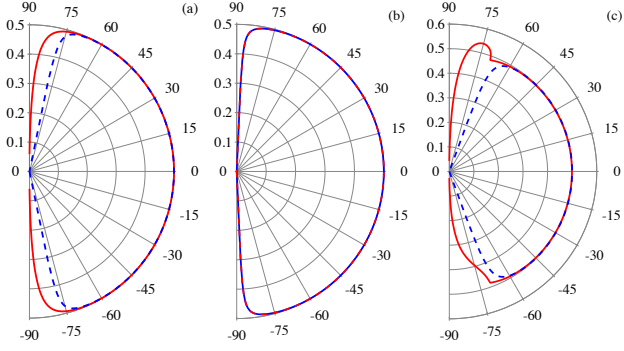}
	\caption{\label{fig:sb:tra} (Color online)
	Angular dependence of the transmission 
	probabilities $T_{+\uparrow}$ (blue dashed line) and $T_{-\uparrow}$ 
	(red solid line) at 
	energy $E=2.5$. The SOC are fixed as follows: 
		(a) $\lambda=0.1$ and $\Delta=0$, 
	(b) $\lambda=0$ and $\Delta=0.1$, and 
	(c) $\lambda=0.5$ and $\Delta=0.1$.
	}
\end{figure}
%
%
%
%
\begin{align}
\Omega_\so(x)  & = \\
& \hspace{-1.5cm}\begin{pmatrix}
\ee^{-\ii \xi_+ -\frac{\theta_+}{2}} & - \ee^{\ii \xi_+ -\frac{\theta_+}{2}} &  
\ee^{-\ii \xi_- -\frac{\theta_-}{2}}& -  \ee^{\ii \xi_- -\frac{\theta_-}{2}} \\
\ee^{\frac{\theta_+}{2}} &   \ee^{\frac{\theta_+}{2}}  & \ee^{\frac{\theta_-}{2}}&  
\ee^{\frac{\theta_-}{2}} \\
\ii \ee^{\frac{\theta_+}{2}}  &  \ii \ee^{\frac{\theta_+}{2}}& -\ii \ee^{\frac{\theta_-}{2}}& - 
\ii \ee^{\frac{\theta_-}{2}} \\
\ii \ee^{\ii \xi_+ -\frac{\theta_+}{2}} & - \ii \ee^{-\ii \xi_+ -\frac{\theta_+}{2}}  & -
\ii \ee^{\ii \xi_- -\frac{\theta_-}{2}} &  \ii \ee^{-\ii \xi_- -\frac{\theta_-}{2}}
\end{pmatrix} \nonumber \\
& \begin{pmatrix}
\mathcal{N}_+ \ee^{\ii k_{x+}+ x} & 0 &  0 & 0 \\
0 &  \mathcal{N}_+  \ee^{-\ii k_{x+} x} & 0 &  0 \\
0 &  0 & \mathcal{N}_-  \ee^{\ii k_{x-} x} & 0 \\
0 & 0 & 0 &  \mathcal{N}_-  \ee^{-\ii k_{x-} x}
\end{pmatrix}\nonumber
\end{align}
%
%
where in the second matrix  the normalization factors
are defined as 
$\mathcal{N}_\alpha= 1/(2\sqrt{v_{\alpha+}\sinh\theta_\alpha})$.

According to  Eq.~(\ref{eq:tran:mat}) the transfer matrix 
for the single interface problem 
is given by the matrix product
$\mathcal{M}=[\Omega_\text{N}(0^-)]^{-1}\Omega_\so(0^+)$. 
From $\mathcal{M}$  we obtain the transmission and the reflection 
probabilities for a spin-up or spin-down incident quasiparticle:
%
%
\begin{align}
T_{+ s} & = \displaystyle\left|\frac{\mathcal{M}_{33}\delta_{s,\uparrow} +
\mathcal{M}_{13}\delta_{s,\downarrow}}{\mathcal{M}_{13}\mathcal{M}_{31}-
\mathcal{M}_{11}\mathcal{M}_{33}}\right|^2 \Upsilon_+(\phi), \\
T_{- s} & = \displaystyle\left|\frac{\mathcal{M}_{31}\delta_{s,\uparrow}+
\mathcal{M}_{11}\delta_{s,\downarrow}}{\mathcal{M}_{13}\mathcal{M}_{31}-
\mathcal{M}_{11}\mathcal{M}_{33}}\right|^2 \Upsilon_-(\phi),\\
R_{\uparrow s} & =  \displaystyle\left|\frac{\mathcal{M}_{31}\mathcal{M}_{23}-
\mathcal{M}_{33}\mathcal{M}_{21}}{\mathcal{M}_{13}\mathcal{M}_{31}-
\mathcal{M}_{11}\mathcal{M}_{33}}\right|^2 \delta_{s,\uparrow}   
\nonumber \\ & \hspace{1.cm} 
+ \left|\frac{\mathcal{M}_{13}\mathcal{M}_{21}-\mathcal{M}_{11}\mathcal{M}_{23}}{\mathcal{M}_{13}\mathcal{M}_{31}-
\mathcal{M}_{11}\mathcal{M}_{33}}\right|^2 \delta_{s,\downarrow}, \\
R_{\downarrow s} & =  \displaystyle \left|\frac{\mathcal{M}_{31}\mathcal{M}_{43}-
\mathcal{M}_{33}\mathcal{M}_{41}}{\mathcal{M}_{13}\mathcal{M}_{31}-\mathcal{M}_{11}
\mathcal{M}_{33}} \right|^2 \delta_{s,\uparrow} \nonumber \\ & \hspace{1cm} 
+\left|\frac{\mathcal{M}_{13}\mathcal{M}_{41}-\mathcal{M}_{11}
\mathcal{M}_{43}}{\mathcal{M}_{13}\mathcal{M}_{31}-\mathcal{M}_{11}
\mathcal{M}_{33}}\right|^2 \delta_{s,\downarrow},
\end{align}
%
%
where $\Upsilon_\alpha(\phi)=\theta(\tilde\phi_\alpha-\phi)\theta(\tilde\phi_\alpha+\phi)$ 
with $\theta(x)$ the Heaviside step function. Here, $T_{\alpha s}$ is the 
probability for an incident quasiparticle with spin projection $s$ to be transmitted
in mode $\alpha$ in the SO region.
Of course, probability current conservation enforces 
$T_{+s}+T_{-s}+R_{\uparrow s}+R_{\downarrow s}=1$.

Figures~\ref{fig:sb:tra} (a)--(c)  show the angular dependence of 
the transmission probabilities for an incident spin-up quasiparticle into 
the $(+)$ and  $(-)$ modes of the SO region for different values of the 
SOCs. 
Panel (a) refers to the case of vanishing intrinsic SOC ($\Delta=0$). 
Here  the $(+)$ and the $(-)$ energy bands are separated by a 
SOC-induced splitting  $\Delta E_\text{ext}=\lambda$. 
Therefore at fixed energy the two propagating modes in the SO region
 have two different momenta, which gives rise to the two different critical angles 
(cf. Eq.~(\ref{eq:crit:angles}) with $\Delta=0$). 
Panel (b) refers to the case $\lambda=0$, where
the SOC opens a gap $\Delta E_\text{int}=2\Delta$ between the
particle- and the hole-branches, however the $(+)/(-)$-modes remain
degenerate. Therefore at fixed 
energy these modes have the same momentum and, as a consequence, 
the same critical angles (cf. Eq.~(\ref{eq:crit:angles}) for $\lambda=0$ and $\Delta\neq0$). 
When both SOCs are finite | the situation illustrated in panel (c) | 
the transmission probabilities exhibit more structure.
For incidence angles smaller than $\tilde \phi_+$
no particular differences with the cases of panels (a) and (b) are visible. 
When the $(+)$ mode is closed, an increase (resp. decrease) of the $(-)$ mode 
transmission is observed for positive (resp. negative) angles, before the 
transmission drops to zero for incidence angles approaching 
$\tilde \phi_-$. The asymmetry between positive and negative 
angles is reversed if the spin state of the incident quasiparticle is reversed.

These symmetry properties can be rationalized by considering the operator 
of mirror symmetry through the $x$-axes.\cite{zhai:2005} 
This consists of the transformation $y\to-y$ and at the same time the inversion of
the spin and the pseudo-spin states. It reads
%
%
\begin{equation}\label{mirror:y}
\mathcal{S}_y = ( \sigma_x \otimes s_y ) R_y ,
\end{equation}
%
%
where $R_y$ transforms $y\to-y$. The operator $\mathcal{S}_y$ commutes with the total Hamiltonian of the system 
$[\mathcal{S}_y, \mathcal{H}_0+\mathcal{H}_\so] =0$, therefore allows for a common basis of eigenstates.  For the scattering states in the SO region (\ref{eq:wf:so}) we have 
$\mathcal{S}_y \chi_+(\xi_+)= \chi_+(\xi_+)$ and 
$\mathcal{S}_y \chi_-(\xi_-)= -\chi_-(\xi_-)$. 
Instead, it induces 
the following transformation on the scattering states (\ref{eq:wf:norm}) 
in the normal region: $\mathcal{S}_y \chi_s(\phi) = \ii \chi_{-s}(-\phi)$. 
By comparing the original scattering matrix with the $\mathcal{S}_y$-transformed 
one we find that 
%
%
\begin{equation}\label{eq:sing:sym}
T_{\alpha,s}(\phi)=T_{\alpha,-s}(-\phi)
\end{equation}
%
%
with $\alpha=\pm$ and $s=\uparrow,\downarrow$, which 
is indeed the symmetry observed in the plots. The asymmetry 
of the transmission coefficients occurs only when
both SOCs are finite.

\subsection{Edge states at the interface}
%
%
\begin{figure}[b]
	\centering
	\includegraphics[width=0.7\columnwidth]{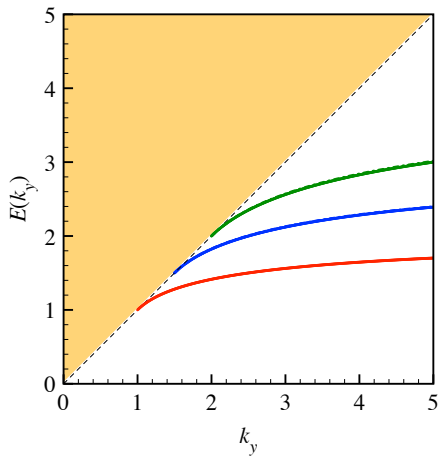}
	\caption{\label{fig:bs} (Color online). Energy dispersion of the edge state at the N-SO 
	interface as a function of the momentum along the interface $k_y$ for different 
	values of SOCs. 
	Solution of the transcendental equation is allowed only for $|k_y|>|E|$ 
	(white area). In all three cases shown $\eta>1/2$: $\Delta=1$ and 
	$\lambda=0.4$ (lower-red line), $\Delta=1.5$ and $\lambda=0.7$  (middle-blue line), 
	and $\Delta=2$ and $\lambda=0.9$ (upper-green line).}
\end{figure}
%
%
In addition to scattering solutions of the DW equation, 
it is interesting to study the possibility that edge states exist 
at the N-SO interface, which propagate {\em along} the interface 
but decay exponentially away from it. 
The interest in these types of solutions is connected to the  study 
of topological insulators. It has been shown | first by Kane and 
Mele\cite{kane:2005} | that  a zig-zag graphene 
nanoribbon with intrinsic SOC supports dissipationless edge states 
within the SOC gap. In fact, similar states are always expected to exist at the 
interface between a topologically trivial and a topologically non-trivial insulator. 
In our case, the latter is represented by graphene with intrinsic SOC. Of course 
SOC-free graphene is not an insulator, however it is topologically trivial and 
edge state solutions do arise for $|k_y|>|E|$. When $E$ is within 
the gap in the SO region the corresponding mode 
is evanescent along the $x$ direction on both sides of the interface.
Note that the edge state we find is different from the one discussed in
Refs.~\onlinecite{kane:2005}, \onlinecite{stauber:2009} where zig-zag or hard-wall
boundary conditions at the edge of the SOC region were imposed.
 
The wave function on the N side then reads
%
%
\begin{align}
\chi_\nn(x) = \begin{pmatrix}
	1 \\
	\frac{-\ii q+\ii k_y}{E}
\end{pmatrix} 
	\left( A \kets{\downarrow}  + B\kets{\uparrow} \right) \ee^{qx}
\end{align}
%
%
with $q=\sqrt{|k_y|^2-E^2}$. 
In the SO region the wave function can be written as
%
%
\begin{align}
\chi_\so(x) & =  
%
C \begin{pmatrix}
	(- q_+ +  k_y)\\
	\ii(E-\Delta)\\
	E-\Delta\\
	\ii(q_+ +  k_y)
\end{pmatrix}  
\ee^{-q_+x} +  D
\begin{pmatrix}
	(q_-  -  k_y)\\
	-\ii(E-\Delta)\\
	E-\Delta\\
	(q_- + \ii k_y)
\end{pmatrix}  
\ee^{-q_-x}\nonumber
\end{align}
%
%
with $q_\alpha=\sqrt{k_y^2-(E-\Delta)(E+\Delta-\alpha \lambda)}$.
The continuity of the wave function at the N-SO interface leads to a 
linear system of equations for the amplitudes $A$ to $D$. The matrix of 
coefficients must have a
vanishing determinant for a non-trivial solution 
to exist. This condition provides a transcendental 
equation for the energy of possible edge states, whose 
solutions are illustrated in Fig.~\ref{fig:bs}\, for different values 
of the intrinsic and extrinsic SOCs. The condition $|k_y|>|E|$ implies 
that solutions only exist outside the shadowed area. 
In addition, they are allowed 
only in the case SOCs open a gap in the energy spectrum, which
occurs when $\eta>1/2$  (see App.~\ref{app:so} and 
Eq.~(\ref{eq:eta})). As can be seen in Fig.~\ref{fig:bs} 
the result is quite insensitive to the precise value of 
the extrinsic SOC.  

Edge states exist only for values of the momentum 
along the interface larger than the intrinsic SOC, \emph{i.e.},
$k_y>k_y^\text{min}=\Delta$. The apparent breaking 
of time-reversal invariance (the dispersion is not even in $k_y$)
is due to the fact that we are considering
a single-valley theory. The full two-valley SOC  Hamiltonian 
is invariant under time-reversal symmetry,
that interchanges the valley quantum number.
This invariance implies that  
solutions for negative values of $k_y$ can be obtained 
by considering the Dirac-Weyl Hamiltonian relative to the other valley.  
The two counter-propagating edge states live then at opposite 
valleys and have opposite spin state
and realize a peculiar 1D electronic system.
%
%
%
\begin{figure}
	\centering
	\includegraphics[width=\columnwidth]{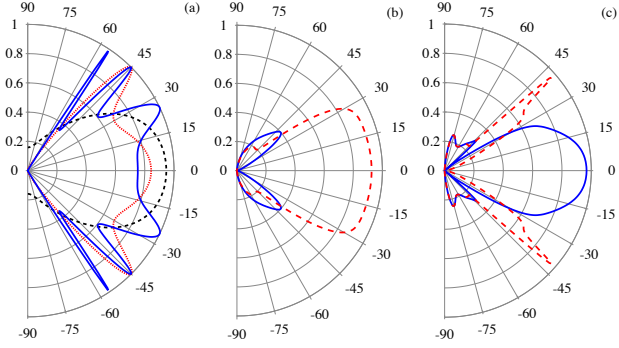}	
	\caption{\label{figure:d}(Color online). Panel (a): Angular plots for $T_{\uparrow\uparrow}$
	as a function of the injection angle for $E=2$, $\Delta=1$ and $\lambda=0$. 
	The three lines correspond to different distance between the interfaces: $d=\pi/2$ 
	(dashed black), $d=\pi$ (dotted red), and $d=2\pi$ (solid blue). The spin-precession 
	length is $\ell_\so=\pi$. When $\lambda=0$  the transmission probability in the spin state
	opposed to the injected spin is always zero. Panel (b) and (c): angular plots of
	$T_{\uparrow\uparrow}$ (solid-blue) and $T_{\downarrow\uparrow}$ (dashed red)
	as a function of the injection angle for $E=2$, $\lambda=1$ and $\Delta=0$. 
	The distance between the two interfaces is $d=\pi$ in panel (a) and $d=2\pi$ in 
	panel (b). The spin-precession length is $\ell_\so=2\pi$.}
\end{figure}
%
%

As mentioned in the Introduction, the intrinsic SOC is estimated to be much 
smaller than the extrinsic one, therefore in a realistic
situation one would not expect the opening of a significant
energy gap and the presence of edge states. It would be 
interesting to explore the possibility to artificially  enhance the 
intrinsic SOC, thereby realizing  the condition  for the occurrence of edge states.


\section{The N-SO-N interface}
\label{NSON}

The analysis of the scattering problem on a N-SO
interface of the previous section can be
straightforwardly  generalized 
to the case of a double N-SO-N interface (SO barrier).
Here the transmission matrix $\mathcal{D}$ 
is given by Eq.~(\ref{eq:tran:mat:2}) with $N=2$. 
The transmission and the reflection probabilities in the case 
of a spin-up or -down incident quasiparticle read
%
%
\begin{align}\label{eq:tf:di}
T_{\uparrow s} & = \displaystyle\left|\frac{\mathcal{D}_{33}\delta_{s,\uparrow} +\mathcal{D}_{13}\delta_{s,\downarrow}}{\mathcal{D}_{13}\mathcal{D}_{31}-\mathcal{D}_{11}\mathcal{D}_{33}}\right|^2, \\
T_{\downarrow s} & = \displaystyle\left|\frac{\mathcal{D}_{31}\delta_{s,\uparrow}+\mathcal{D}_{11}\delta_{s,\downarrow}}{\mathcal{D}_{13}\mathcal{D}_{31}-\mathcal{D}_{11}\mathcal{D}_{33}}\right|^2, \\
R_{\uparrow s} & =  \displaystyle\left|\frac{\mathcal{D}_{31}\mathcal{D}_{23}-\mathcal{D}_{33}\mathcal{D}_{21}}{\mathcal{D}_{13}\mathcal{D}_{31}-\mathcal{D}_{11}\mathcal{D}_{33}}\right|^2 \delta_{s,\uparrow}  
 \nonumber \\ & \hspace{1.cm} 
+ \left|\frac{\mathcal{D}_{13}\mathcal{D}_{21}-\mathcal{D}_{11}\mathcal{D}_{23}}{\mathcal{D}_{13}\mathcal{D}_{31}-\mathcal{D}_{11}\mathcal{D}_{33}}\right|^2 \delta_{s,\downarrow}, \\
R_{\downarrow s} & =  \displaystyle \left|\frac{\mathcal{D}_{31}\mathcal{D}_{43}-\mathcal{D}_{33}\mathcal{D}_{41}}{\mathcal{D}_{13}\mathcal{D}_{31}-\mathcal{D}_{11}\mathcal{D}_{33}} \right|^2 \delta_{s,\uparrow}
 \nonumber \\ & \hspace{1cm} 
+ \left|\frac{\mathcal{D}_{13}\mathcal{D}_{41}-\mathcal{D}_{11}\mathcal{D}_{43}}{\mathcal{D}_{13}\mathcal{D}_{31}-\mathcal{D}_{11}\mathcal{D}_{33}}\right|^2 \delta_{s,\downarrow}\, .
\end{align}
%
%

In this case there is an additional parameter which controls the scattering 
properties of the structure,  namely the width $d$ of the SO region. In order 
to compare this length to a characteristic length scale of the system, we 
introduce the spin-precession length defined as
%
\begin{equation}\label{eq:spin:pre:len}
\ell_\so = 2\pi \frac{\hbar v_\text{F}}{\lambda+2\Delta}\,.
\end{equation}
%
%
The intrinsic SOC alone  cannot induce a spin precession 
on the carriers injected into the SO barrier | an injected spin state, 
say up, is obviously never converted into a spin-down state. 
Figu\-re~\ref{figure:d}(a) shows the angular dependence of 
the transmission in the case of injection of spin-up. The behavior
of the transmission as a function of the injection angle depends 
sensitively on the width $d$ compared to the spin-precession length. 
For small width $d<\ell_\so$ (dashed line) the 
transmission is a smooth decreasing function of $\phi$ and
stays finite also for $\phi$ larger than the critical angle. 
In the case $d\ge\ell_\so$ (dotted- and solid-lines) instead
the transmission probability exhibits a resonant behavior and
drops to zero as soon as the injection angle equals the critical angle.
%
%
\begin{figure}
	\centering
	\includegraphics[width=\columnwidth]{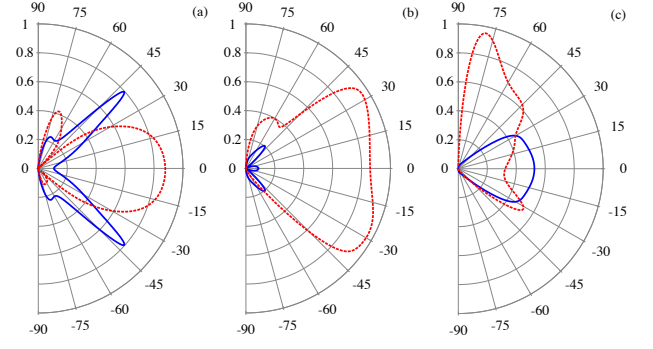}
	\caption{\label{figure:f}(Color online). Angular plot of $T_{\uparrow\uparrow}$ 
	(solid-blue) and  $T_{\downarrow\uparrow}$ (dashed-red) 
	as a function of the injection angle for $E=2$, $\lambda=1$ and 
	(a) $\Delta=\lambda/4$, (b) $\Delta=\lambda/2$, and $\Delta=\lambda$. 
	The distance between the two interfaces is kept fixed to $d=\ell_\so$.}
\end{figure}
%
%

When only the extrinsic SOC is finite, 
the transmission behavior changes drastically. 
Two different critical angles appear | the biggest coincides 
usually with $\pi/2$. 
The extrinsic SOC induces spin precession because of 
the coupling between the pseudo- and the real-spin.
This is illustrated  in Fig.~\ref{figure:d}(b)-(c). 
In Panel (b)  we consider the case of spin-up injection 
with $d=\ell_\so/2$. At normal incidence  the transmission 
is entirely in the spin-down channel (dashed line). 
Moving away from normal incidence, the transmission in the 
spin-up channel (solid line) increases from zero and, 
after the first critical angle, the transmissions in spin-up and spin-down 
channels tend to coincide. 
In panel (c) the width of the barrier is set to $d=\ell_\so$. Here, 
the width of the SO region permits to an injected carrier  at normal 
incidence to perform a complete precession of  its spin state 
| the transmission is in the spin-up channel. For finite injection angles 
the spin-down transmission (dashed line) also becomes finite. 
For $\phi\lesssim\tilde\phi_+$ the transmission in the spin-up channel 
is almost fully suppressed while that in the spin-down channel is large. 
Finally, for $\phi>\tilde\phi_+$ the two transmission coefficients 
do not show appreciable difference.

%
%
\begin{figure}
	\centering
	\includegraphics[width=0.7\columnwidth]{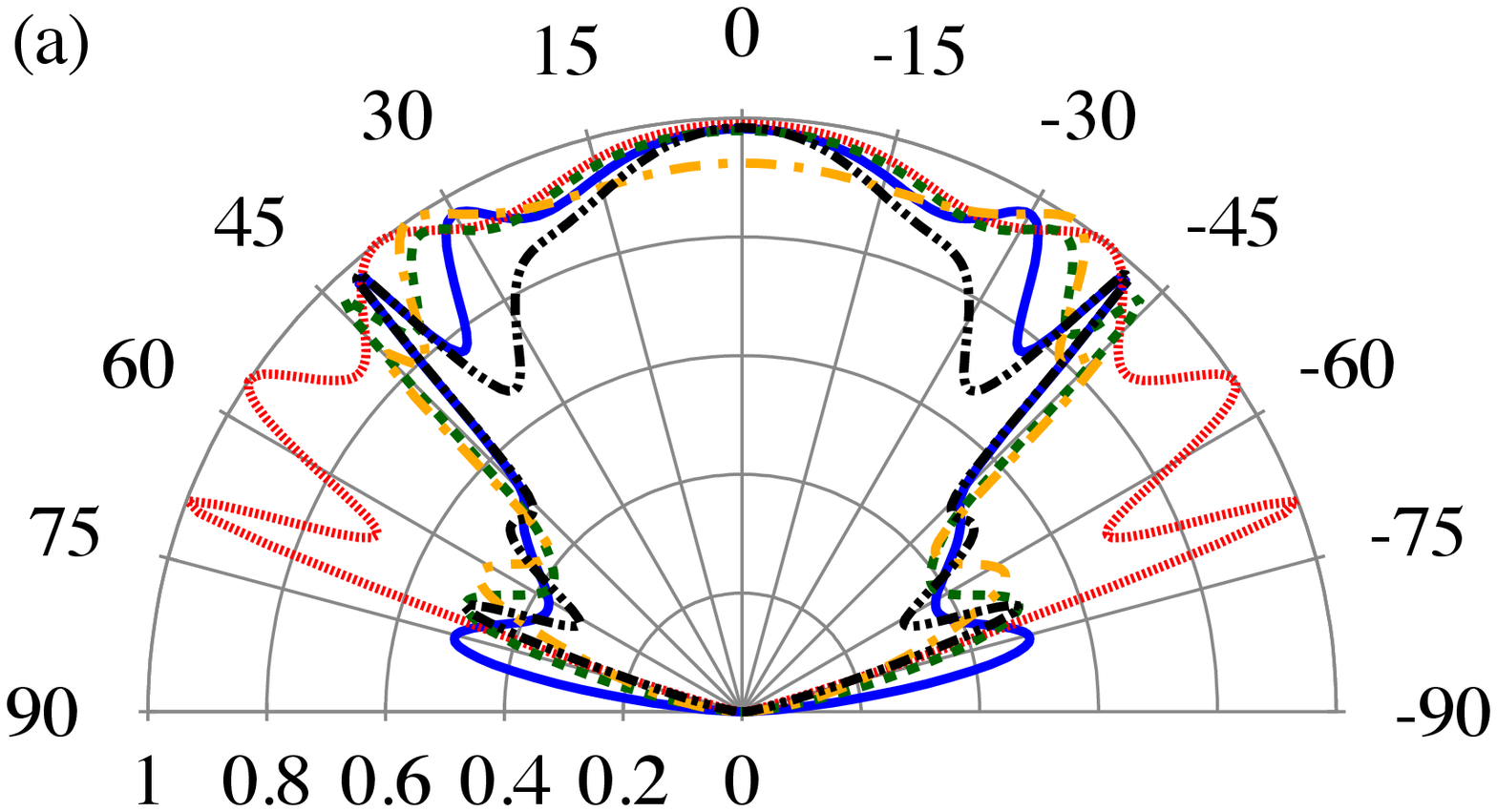}
	\includegraphics[width=0.6\columnwidth]{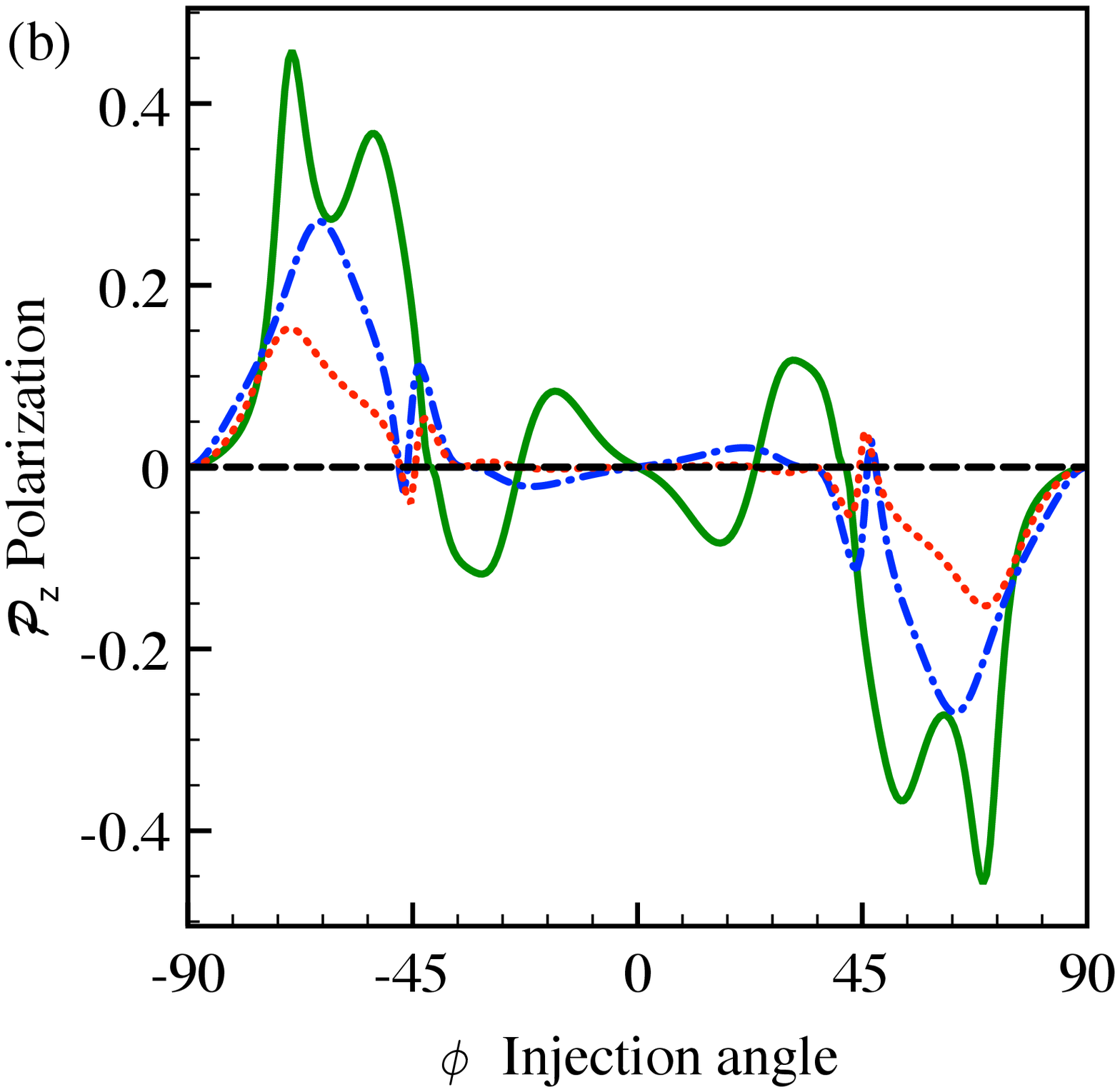}
	\caption{\label{figure:g} (Color online). Panel (a): total transmission $T$ as a 
	function of the injection angle for $E=2$, $d=2\pi$ and several values of SOCs: 
	$\lambda=1$ and $\Delta=0$ (blue-solid line), 
	$\lambda=0$ and $\Delta=0.5$ (red-dotted line), 
	$\lambda=1$ and $\Delta=\lambda/4$  (yellow-dashed line), 
	$\Delta=\lambda/2$ (orange-dashed-dotted line), and $\lambda=\Delta$
	 (black-dotted-dotted-dashed line). Panel (b): $z$-component of the spin 
	 polarization ${\mathcal P}_z$ as a function on the injection angle 
	 for $E=2$ and $d=2\pi$ and the following values of the SOCs:
	 $\lambda=1$, $\Delta=0$ and $\lambda=0$, $\Delta=1$ (same black-dashed line), 
	 $\lambda=1$ and $\Delta=\lambda/4$ (red-dotted), 
	 $\Delta=\lambda/2$ 
	 (blue-dotted-dashed line), and $\Delta=\lambda$ (green-solid line).}
\end{figure}
%
%

In the case where both extrinsic and intrinsic SOC are finite, 
the transmission probability exhibits a richer structure.
We focus again on the case of injection of spin-up quasiparticles. 
Moreover we fix the width of the SO region so that it is always 
equal to the spin-precession length $d=\ell_\so$. 
Fig.~\ref{figure:f}  illustrates the transmission probabilities $T_{s\uparrow}$
for three values of the ratio $\Delta/\lambda=1/4,1/2,1$. 
Notice that from panel (a) to (c) the width of SO region decreases.

The symmetry properties of the transmission function
can be rationalized by using the symmetry operation~(\ref{mirror:y}).
Proceeding in a similar manner as in the  case of the single interface,  
for the SO barrier we find the following symmetry relations 
%
%
\begin{subequations}\label{eq:sym:db}
\begin{align}
T_{s,s}(\phi) & = T_{s,s}(-\phi)\,, \\
T_{s,-s}(\phi) & = T_{-s,s}(-\phi)\, ,
\end{align}
\end{subequations}
%
%
which are confirmed by the explicit calculations.

%
%

So far we have considered the injection of a pure 
spin state |  the injected carrier was either in a spin-up state or a 
spin-down state. Following Ref. \onlinecite{marigliano2}
we now address the transmission of an unpolarized 
statistical mixture of spin-up and spin-down carriers. 
This will characterize the spin-filtering properties of the SO barrier. 
In the injection N region, an unpolarized statistical mixture of spins is 
defined by the density matrix
%
%
\begin{equation}\label{eq:rho:in}
\rho_\text{in} = \frac{1}{2} \ket{\chi_\uparrow}\bra{\chi_\uparrow} +
\frac{1}{2} \ket{\chi_\downarrow}\bra{\chi_\downarrow} ,
\end{equation}
%
%
where $\ket{\chi_s}\equiv \ket{s}\otimes \ket{\sigma}$ with 
$\ket{\sigma}= (1/\sqrt{2})(1,\ee^{\ii \phi})$ corresponds to a 
pure spin state. When traveling  through the SO region, 
the injected spin-unpolarized state is subjected to
spin-precession. The density matrix in the output N region can 
be expressed in terms of the transmission functions~(\ref{eq:tf:di}) as
%
%
\begin{equation}\label{eq:rho:out}
\rho_\text{out} = \frac{1}{2} T_\uparrow \ket{\zeta_\uparrow}\bra{\zeta_\uparrow} + \frac{1}{2} T_\downarrow \ket{\zeta_\downarrow}\bra{\zeta_\downarrow},
\end{equation}
%
%
where the coefficients $T_s=T_{\uparrow s}+T_{\downarrow s}$ are the 
total transmissions for fixed injection state. The spinor part is defined as
%
%
\begin{equation}\label{eq:spinor:sm}
\ket{\zeta_s}=\frac{1}{\sqrt{T_s}} 
\begin{pmatrix}
t_{\uparrow s} \\ t_{\downarrow s}
\end{pmatrix} \otimes \ket{\sigma},
\end{equation}
%
%
where $t_{s',s}$ are the transmission amplitudes for incoming
(resp. outgoing) spin $s$ (resp. $s'$). The output density matrix 
is used to define the total transmission 
%
%
\begin{equation}\label{eq:tran:unp}
T=\frac{T_\uparrow +T_\downarrow}{2}
\end{equation}
%
%
and the expectation value of the $z$ component of the spin-polarization
%
%
\begin{equation}\label{eq:pol:z}
\mathcal{P}_z = \frac{1}{2} \left( T_{\uparrow\uparrow} + 
T_{\uparrow\downarrow} - T_{\downarrow\uparrow} - 
T_{\downarrow\downarrow}\right)\,. 
\end{equation}
%
%

In Fig.~\ref{figure:g} we report the total 
transmission (panel~(a)) and the $z$-component of the 
spin-polarization (panel~(b)) as a function of the injection 
angle for fixed energy and width of the SO region. 
We observe that for an unpolarized injected state
the transmission probability is an even function of the injection 
angle $T(\phi)=T(-\phi)$. Moreover, for injection angles 
larger than the first critical angle $\phi > \tilde\phi_+$, the transmission 
has an upper bound at $T = 1/2$. On the contrary $\mathcal{P}_z$ is 
an odd function of the injection angle $\mathcal{P}_z(\phi)=-\mathcal{P}_z(-\phi)$. 
It is zero when at least one SOC is zero. 
When both SOC parameters are finite $\mathcal{P}_z$ 
is finite and reaches the largest values for 
$\phi > \tilde\phi_+$. The maxima in this case increase 
as a function of the intrinsic SOC.

To experimentally observe this polarization effect 
the measurement should not involve an average over
the angle $\phi$, which, otherwise | due to the antisymmetry 
of $\mathcal{P}_z$ | would wash out the effect.  To achieve this, one could 
use, {\it e.g.}, magnetic barriers,\cite{ale,luca1} 
which are known to act as wave vector filters.

%
%
\section{Conclusions}
\label{sec:5}

In this paper we have studied the spin-resolved transmission
through SO nanostructures in graphene, {\em i.e.}, systems where  
the strength of SOCs | both intrinsic and extrinsic | is spatially modulated.
We have considered the case of an interface separating a normal region from
a SO region, and a barrier geometry with a region of finite SOC 
sandwiched between two normal regions.  
We have shown that | because of the lift of spin degeneracy due to the SOCs
| the scattering at the single interface gives rise to 
spin-double refraction: a carrier injected from the normal region 
propagates into the SO region along two different directions
as a superposition of the two available channels. 
The transmission into each of the two channels depends sensitively 
on  the injection angle and on the values of SOC parameters. 
In the case of a SO barrier, this result can be used to select preferential 
directions  along which the spin polarization of an initially unpolarized carrier 
is strongly enhanced. 

We have also analyzed the edge states occurring in the single  
interface problem in an appropriate range of
parameters. These states exist when the SOCs
open a gap in the energy spectrum and
correspond to the gapless edge states supported by the 
boundary of topological insulators.

A natural follow-up to this work would be 
the detailed analysis of transport properties of such SO
nanostructures. From our results for 
the transmission probabilities, spin-resolved conductance
and noise could easily be calculated by means of the 
Landauer-B\"uttiker formalism. 
Moreover we plan to study other geometries, 
as, \emph{e.g.}, nanostructures
 with a periodic modulation of SOCs. 
The effects of various types of impurities on the properties 
discussed here is yet another interesting issue to address.

We hope that our work will stimulate further theoretical 
and experimental investigations on spin transport properties 
in graphene nanostructures.

\acknowledgments
We gratefully acknowledge helpful discussions with  L.~Dell'Anna,
R.~Egger, H.~Grabert, M.~Grifoni, W. H\"ausler,
V.~M.~Ramaglia, P.~Recher and D.~F.~Urban. 
The work of DB is supported by the Excellence Initiative of the German 
Federal and State Governments. The work of ADM is supported by the SFB/TR 12 of the DFG.

\appendix

\section{Graphene with uniform spin-orbit interactions.\label{app:so}}

%
%
\begin{figure}[b]
	\begin{center}
	\includegraphics[width=0.9\columnwidth]{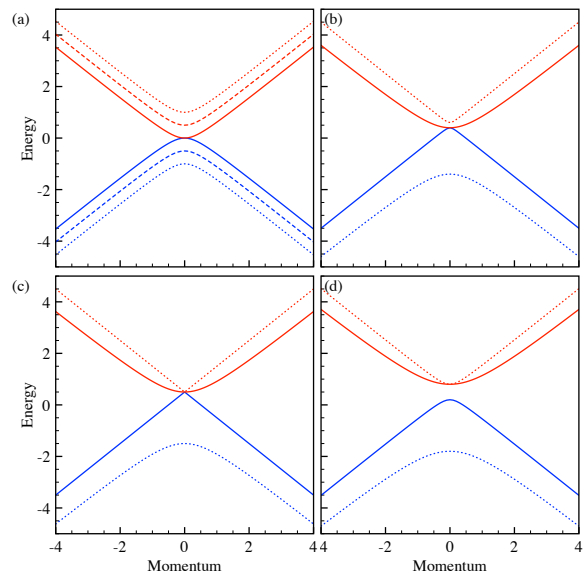}
	\caption{\label{fig:spec} Spectrum of the DW Hamiltonian with
	intrinsic and Rashba SOC as a function of $k_x$ for $k_y=0$ . 
	Panel (a): dashed lines refer to $\Delta=0.5$ and $\lambda=0$;
	solid and dotted lines refer to $\Delta=0$ and $\lambda=1$. 
	Panel (b): $\Delta=0.4$ and $\lambda=1$. 
	Panel (c): $\Delta=0.5$ and $\lambda=1$. 
	Panel (d): $\Delta=0.8$ and $\lambda=1$.}
	\end{center}
\end{figure}
%
%

In this appendix we briefly review the basic properties of
DW fermions in graphene with homogeneous  
SO interactions.\cite{rashbagraphene}  
The energy eigenstates are
plane waves  $\psi \sim \Phi({\bf k})\ee^{\ii \mathbf{k}\cdot\mathbf{r}}$
with $\Phi$ a four-component spinor and
eigenvalues given by ($v_\text{F}=\hbar = 1$)
%
%
\begin{equation}\label{eq:so:spectrum}
E_{\alpha,\epsilon}(\mathbf{k}) =\frac{\alpha\lambda}{2} +\epsilon \sqrt{k_x^2+k_y^2 + 
\left(\Delta-\frac{\alpha\lambda}{2} \right)^2},
\end{equation}
%
%
where $\alpha=\pm$ and $\epsilon=\pm$.
The energy dispersion as a function of $k_x$ 
at fixed $k_y=0$ is illustrated in Fig.~\ref{fig:spec} for several values 
of $\Delta$ and $\lambda$. 
The index $\epsilon=\pm$ specifies the particle/hole branches 
of the spectrum. 
The eigenspinors $\Phi_{\alpha,\epsilon}({\bf k})$ read
%
%
\begin{align}
\Phi^\text{T}_{\alpha,\epsilon} (\mathbf{k}) =&
\frac{1}{2\sqrt{\cosh \theta_\alpha}} 
\times\\ & 
(\ee^{-\ii\phi-\epsilon \theta_\alpha/2},
\epsilon \ee^{\epsilon\theta_\alpha/2},\ii\alpha\epsilon \ee^{\epsilon\theta_\alpha/2},\ii\alpha
\ee^{\ii\phi-\epsilon\theta_\alpha/2}), \nonumber
\end{align}
%
%
where ${}^\text{T}$ denotes transposition and
%
%
\begin{align}
&\sinh \theta_\alpha = \frac{\alpha\lambda/2 -\Delta}{k},\\
& \ee^{\ii\phi} = \frac{k_x+\ii k_y}{k}, \label{phi}
\end{align}
%
%
with $k=\sqrt{k_x^2+k_y^2}$.  
The spin operator components are expressed as 
$S_j=\frac{1}{2}s_j\otimes \sigma_0$. 
Their expectation values in the 
eigenstate $\Phi_{\alpha,\epsilon}$ read
%
%
\begin{subequations}\label{eq:spin:state}
\begin{align}
\langle S_x \rangle &= \frac{-\epsilon \alpha \sin \phi}{2\cosh \theta_\alpha},  \\
\langle S_y \rangle &= \frac{\epsilon \alpha \cos \phi}{2\cosh \theta_\alpha}, \\
\langle S_z \rangle &= 0,
\end{align}
\end{subequations}
%
%
which shows that the product $\epsilon \alpha$ coincides with 
the sign of the expectation value of the spin projection
along the in plane direction perpendicular to the direction of propagation.
For vanishing extrinsic SOC, the eigenstates $\Phi_{\alpha,\epsilon}$
reduce to linear combinations of eigenstates of $S_z$.

Similarly, the expectation value of the pseudo-spin operator $\bm \sigma$
is given by
%
%
\begin{subequations}
\begin{align}
\langle \sigma_x \rangle &= \frac{\epsilon \cos \phi}{\cosh \theta_\alpha}, \label{psx}\\
\langle \sigma_y \rangle &= \frac{\epsilon \sin \phi}{\cosh \theta_\alpha}. \label{psy}
\end{align}
\end{subequations}
%
%
Since the SOCs in graphene do not depend on momentum,
the velocity operator still coincides with the pseudo-spin operator:
${\bf v}=\dot {\bf r} = \ii[\mathcal{H},{\bf r} ]={\bm  \sigma}$. Thus 
the velocity expectation value in the state $\Phi_{\alpha,\epsilon}$
is given by Eqs. (\ref{psx}--\ref{psy}). Alternatively, it can be obtained 
from the energy dispersion as
%
%
\begin{align}
& \mathbf{v}_{\alpha,\epsilon} = \nabla_\mathbf{k}
E_{\alpha,\epsilon}
=  \frac{\epsilon \mathbf{k}}{\sqrt{k^2 +\left(\Delta-\frac{\alpha \lambda}{2}\right)^2}}\,. \label{eq:so:velocityx}
\end{align}
%
%
The group velocity is then independent of the modulus of the
wave vector if either the SOCs vanish or $\Delta=\alpha \lambda/2$.


\begin{thebibliography}{99}

\bibitem{geim}
K.~S. Novoselov, A.~K. Geim, S.~V. Morozov, D. Jiang, Y. Zhang, S.~V. Dubonos, 
I.~V. Griegorieva, and A.~A. Firsov, 
Science {\bf 306}, 666 (2004); Nature (London) {\bf 438}, 197 (2005).

\bibitem{kim} 
Y. Zhang, Y.~W. Tan, H.~L. Stormer, and P. Kim,
Nature (London) {\bf 438}, 201 (2005).

\bibitem{reviews} 
For recent reviews, see A.~K. Geim and K.~S. Novoselov,
Nature Mat. {\bf 6}, 183 (2007); A.H. Castro Neto, F. Guinea, 
N.~M.~R. Peres, K.~S. Novoselov, and A.~K. Geim, 
Rev. Mod. Phys. {\bf 81}, 109 (2009); A.~K. Geim, Science {\bf 324}, 1530 (2009).

\bibitem{beenakker:2008} 
C.~W.~J. Beenakker, Rev. Mod. Phys. \textbf{80}, 1337 (2008).

\bibitem{semenoff} 
G.~W. Semenoff,
Phys. Rev. Lett. {\bf 53}, 2449 (1984).

\bibitem{divincenzomele}
D.~P. Di Vincenzo and E.J. Mele,
Phys. Rev. B {\bf 29}, 1685 (1984).

\bibitem{spintronics}I. \v Zuti\'c, J. Fabian, S. Das Sarma,
Rev. Mod. Phys. {\bf 76}, 323 (2004).

\bibitem{Trauzettel:2008} B. Trauzettel, D.V. Bulaev, D. Loss, and G. Burkard,
Nature Phys. \textbf{3}, 192 (2007).

\bibitem{expspin1}
E.~W. Hill, A.K.Geim, K.~S. Novoselov, 
F. Schedin, and P. Blake,
IEEE Trans. Magn. {\bf 42}(10),2694 (2006).

\bibitem{expspin2}
N. Tombros, C. Jozsa, M. Popinciuc, H.~T. Jonkman, and B.~J. van Wees,
Nature {\bf 448}, 571 (2007).

\bibitem{expspin3}
S. Cho, Y.F. Chen, and M.~S. Fuhrer,
Appl. Phys. Lett. {\bf 91} 123105 (2007).

\bibitem{expspin4} M. Nishioka and A.~M. Goldman,
Appl. Phys. Lett. {\bf 90} 252505 (2007).

\bibitem{expspin5}
C. J\'ozsa, M. Popinciuc, N. Tombros, H.~T. Jonkman, and B.~J. van Wees,
Phys. Rev. Lett. {\bf 100}, 236603 (2008)

\bibitem{expspin6}
N. Tombros, S. Tanabe, A. Veligura, C. Jozsa, M. Popinciuc, H.~T. Jonkman, and
B.~J. van Wees,
Phys. Rev. Lett. {\bf 101}, 046601 (2008).

\bibitem{kane:2005} 
C.~L. Kane and E.~J. Mele, 
Phys. Rev. Lett. {\bf 95}, 226801 (2005).

\bibitem{soguinea}
D. Huertas-Hernando, F. Guinea, and A. Brataas,
Phys. Rev B {\bf 74} 155426 (2006).

\bibitem{so2}
Hongki Min, J.E. Hill, N.~A. Sinitsyn, B.~R. Sahu, L. Kleinman, and
A.H. MacDonald,
Phys. Rev. B {\bf 74}, 165310 (2006).

\bibitem{so3}
Y. Yao, F. Ye, X.~L. Qi, S.~C. Zhang, and Z. Fang,
Phys. Rev. B {\bf 75}, 041401(R) (2007).

\bibitem{so4} 
J.~C. Boettger and S.~B. Trickey, Phys. Rev. B \textbf{75}, 121402(R)
(2007); Phys. Rev. B \textbf{75}, 199903(E) (2007).

\bibitem{so5} 
M. Zarea and N. Sandler, Phys. Rev. B \textbf{79}, 165442 (2009).

\bibitem{rashbagraphene} 
E.~I. Rashba, 
Phys. Rev. B {\bf 79}, 161409(R) (2009).

\bibitem{guinea2009} D. Huertas-Hernando, F. Guinea, and A. Brataas,
Phys. Rev. Lett. {\bf 103}, 146801 (2009).

\bibitem{ertler:2009} C. Ertler, S. Konschuh, M. Gmitra, and J. Fabian, 
Phys. Rev. B \textbf{80}, 041405(R) (2009).

\bibitem{philip}
P. Ingenhoven, J.~Z. Bern\'ad, U. Z\"ulicke, and R. Egger
Phys. Rev. B {\bf 81}, 035421 (2010) .

\bibitem{varykhalov:2008} 
A. Varykhalov, J. S\'anchez-Barriga, A.~M. Shikin, C. Biswas, E. Vescovo,
A. Rybkin, D. Marchenko, and O. Rader, 
Phys. Rev. Lett. \textbf{101}, 157601 (2008).

\bibitem{rader:2009} 
O. Rader, A. Varykhalov, J. S\'anchez-Barriga, D. Marchenko, A. Rybkin, and A.~M. Shikin, 
Phys. Rev. Lett. \textbf{102}, 057602 (2009).

\bibitem{varykhalov:2009} A. Varykhalov and O. Rader, 
Phys. Rev. B \textbf{80}, 035437 (2009).

\bibitem{marigliano1}
V.~M. Ramaglia, D. Bercioux, V. Cataudella, G. De Filippis, 
C.~A. Perroni, and F. Ventriglia, 
Eur. Phys. J. B {\bf 36}, 365 (2003).

\bibitem{khodas} 
M. Khodas, A. Shekhter, and A.~M. Finkel'stein, 
Phys. Rev. Lett. {\bf 92}, 086602 (2004).

\bibitem{marigliano2}
V.~M. Ramaglia, D. Bercioux, V. Cataudella, G. De Filippis, 
and C.~A. Perroni, J. Phys.: Condens. Matter {\bf 16}, 9143 (2004).

\bibitem{rashba:1960} E.~I. Rashba, Fiz. Tverd. Tela 
(Leningrad) \textbf{2}, 1224 (1960) [Sov. Phys. Solid State \textbf{2}, 1109 (1960)].

\bibitem{stauber:2009} 
T. Stauber and J. Schliemann, New J. Phys. \textbf{11}, 115003 (2009).

\bibitem{yamakage:2009} 
A. Yamakage, K.-I. Imura, J. Cayssol, and Y. Kuramoto, 
EPL {\bf 87}, 47005 (2009).

\bibitem{guinea:2009} F. Guinea, M.~I. Katsnelson, and A.~G. Geim, 
Nature Phys. \textbf{6}, 30 (2009). 

\bibitem{mckellar:1987} 
B.~H.~J. McKellar and G.~J. Stephenson, Jr.,  Phys. Rev. C \textbf{35}, 2262 (1987).

\bibitem{peeters}
M. Barbier, F.~M. Peeters, P. Vasilopoulos, and J.~M. Pereira ,
Phys. Rev. B {\bf 77}, 115446.

\bibitem{luca1}
L. Dell'Anna and  A. De Martino, 
Phys. Rev. B {\bf 79}, 045420 (2009); Phys. Rev. B {\bf 80}, 089901(E) (2009).

\bibitem{luca2}
L. Dell'Anna and  A. De Martino, 
Phys. Rev. B {\bf 80}, 155416 (2009).

\bibitem{footnote1}
Eigenspinors are normalized in order to ensure 
probability flux conservation across the interface.

\bibitem{footnote2}
For incidence angle larger than one of the two critical angles, 
the respective refraction angle becomes complex:
%
\begin{equation*}
\xi_\alpha= \frac{\pi}{2}-\ii \xi'_\alpha\,,
\end{equation*}
%
where the correct determination of the imaginary part is 
obtained for $\xi'>0$ by the relations
$\sin\xi_\alpha  = \cosh \xi'_\alpha$ and
$\cos \xi_\alpha  = \ii \sinh \xi'_\alpha$.

\bibitem{zhai:2005} F. Zhai and H.~Q. Xu, Phys. Rev. Lett. \textbf{94}, 246601 (2005).

\bibitem{ale}  A. De Martino, L. Dell'Anna, and R. Egger,
Phys. Rev. Lett. {\bf 98}, 066802 (2007);
Sol. St. Comm. {\bf 144}, 547 (2007).

\end{thebibliography}
\end{document}